# Graphene morphology regulated by nanowires patterned in parallel on a substrate surface

Zhao Zhang, Teng Li[*]

*Department of Mechanical Engineering, University of Maryland, College Park, Maryland 20742, USA*

**Abstract**

The graphene morphology regulated by nanowires patterned in parallel on a substrate surface is quantitatively determined using energy minimization. The regulated graphene morphology is shown to be governed by the nanowire diameter, the nanowire spacing and the interfacial bonding energies between the graphene and the underlying nanowires and substrate. We demonstrate two representative regulated graphene morphologies and determine critical values of the nanowire spacing, nanowire diameter and interfacial bonding energies at which graphene switches between the two representative morphologies. Interestingly, we identify a rule-of-thumb formula that correlates the critical nanowire spacing, the critical interfacial bonding energies and the nanowire diameter in quite well agreement with the full-scale simulation results. Results from the present study offer guidelines in nano-structural design to achieve desired graphene morphology via regulation with a resolution approaching the atomic feature size of graphene.

[*] Author to whom correspondence should be addressed. Electronic mail: LiT@umd.edu. Tel.: 301-405-0364. FAX: 301-314-9477.

## I. INTRODUCTION

Graphene, a monolayer of graphite, has rapidly emerged as a rising star of materials science and condensed-matter physics,[1-3] largely due to its exceptional electronic, mechanical, and thermal and optical properties.[1-7] These extraordinary properties of graphene have also sparked a surge of scientific and technological interest in graphene-based electronics,[8-11], driven by the desire to overcome the fast-approaching fundamental limits of silicon in 15~20 years.[12] Despite of the promising future of graphene-based electronic devices, there are still significant challenges to their realization, largely due to the difficulty of precisely controlling the graphene properties. Graphene is intrinsically non-flat and corrugates randomly.[13,14] Since the corrugating physics of atomically-thin graphene is strongly tied to its electronics properties,[4,15-17] these random corrugations lead to unpredictable graphene properties, which are fatal for nano-electronic devices.

Recent experimental observations and computational simulations shed lights on new pathways to achieve fine control of graphene properties.[18-23] For example, recent experiments show that the intrinsic random corrugations in graphene can be suppressed by the underlying atomically smooth substrate surfaces.[20] Monolayer and few-layer graphene fabricated via mechanical exfoliation are shown to partially follow the surface morphology of the underlying substrates.[18,21] The resulting graphene morphology is *regulated*, distinct from the intrinsic random corrugations in freestanding graphene. The substrate-regulated graphene morphology results from the interplay between the graphene-substrate interfacial bonding energy and the strain energy of the graphene-substrate system (to be further detailed in the next section). These experimental observations have motivated recent computational studies on graphene morphology regulated by underlying substrates with nanoscale engineered surface patterns (e.g., surface grooves,[22] herringbone and checkerboard wrinkles[23]). While it is rather challenging to control the graphene morphology via direct manipulation of individual carbon atoms,[24] patterned substrate surfaces with nanoscale resolution are readily achievable through existing nano-fabrication techniques[25] (e.g., nanoimprint lithography, self-assembly). The regulated morphology of graphene on such patterned substrate surfaces potentially allows fine control of the electronic properties of graphene.

The abovementioned recent computational studies focused on graphene morphology regulated by substrate surfaces with feature size on the order of ten nanometers. To further explore the abundant opportunities of fine tuning graphene morphology via surface/interface regulation, in this paper we further extend our earlier energetic framework[22] to study the graphene morphology regulated by nanowires patterned on a substrate surface. The past decade has seen significant progresses in fabricating low-dimensional nanostructures[26,27] (e.g., nanowires, nanoparticles) with controllable size and shape. For example, silicon nanowires with diameter of *one nanometer* have been demonstrated.[28] There has also been promising demonstration of controllable patterning of nanowires and nanoparticles on substrate surface via self-assembly[29] or epitaxial growth.[30] Nanowires with diameters of down to one nanometer patterned on substrate surface offer new platforms to regulate graphene morphology with a resolution approaching the atomic feature size of graphene. Furthermore, existing computational studies considered the graphene fully or partially conforming to idealized substrate surface features (e.g., sinusoidal grooves, herringbone corrugations), which thus results in well defined graphene morphology (i.e., similar to that of the substrate surface but with different out-of-surface amplitude). An array of nanowires patterned on a substrate defines a non-smooth surface feature on which the graphene morphology can be fine tuned. The resulting graphene morphology cannot be readily predicted or extrapolated from the results of the abovementioned studies. Moreover, the energetic framework in earlier studies did not include the contribution of stretching to the graphene strain energy. Aiming to address the above concerns and establish effective strategies for precise extrinsic regulation of the graphene morphology, here we quantitatively determine the graphene morphology regulated by nanowires evenly patterned in parallel on a substrate surface (Fig. 1a). The rest of the paper is organized as follows. Section II describes the energetic of graphene morphology regulated by nanowires patterned on a substrate surface; Section III lays out the computational model; Section IV describes the simulation results of two cases: (1) graphene regulated by widely spaced nanowires on a substrate surface, and (2) graphene regulated by densely spaced nanowires on a substrate surface, from which a coherent understanding of graphene morphology regulated by patterned nanowires on a substrate surface is obtained. Particular efforts are focused on the effects of nanowire spacing and interfacial bonding energy on the regulated graphene morphology, as well as an unexpected snap-through instability of the graphene morphology; Section V includes the concluding remarks.

## II. ENERGETIC OF GRAPHENE MORPHOLOGY REGULATED BY NANOWIRES PATTERNED ON A SUBSTRATE SURFACE

For graphene fabricated via mechanical exfoliation on a substrate surface with patterned nanowires (or transfer printed[9] from a transfer substrate surface to a substrate surface with patterned nanowires), the graphene–substrate and graphene-nanowires interfacial bonding energies are usually weak (e.g., characterized by van der Waals interactions). As the graphene corrugates to follow the surface envelope of the substrate with patterned nanowires (e.g., Fig. 1a), the interaction energies (i.e., between graphene and substrate, and between graphene and nanowires) decrease due to the nature of van der Waals interaction; on the other hand, the conforming corrugations of the graphene result in the increase of the system strain energy due to the intrinsic bending rigidity of graphene (an uncorrugated graphene has zero strain energy). At equilibrium, the sum of the total interaction energies and the system strain energy reaches its minimum, from which the morphology of graphene regulated by the patterned nanowires on the substrate surface can be determined.

Figure 1 depicts the model configuration considered in this paper, in which the morphology of a blanket graphene is regulated by an array of nanowires of diameter $d_{nw}$ equally spaced in parallel on a substrate surface. Given the periodicity of the structure, in the rest of the paper we focus on one period of the graphene and the underlying nanowire and substrate (Fig. 1a). When the spacing between the nanowires, $L_{nw}$, is large, the graphene tends to wrap around each individual nanowire (e.g., Fig. 1b). The regulated graphene morphology in one period can be divided into two regions. In region I, graphene corrugates to wrap around the nanowire; In region II the graphene remains flat on the substrate surface. The maximum amplitude of the graphene corrugation in region I is assumed to be equal to the diameter of the nanowire, as suggested in a previous study of the graphene morphology regulated by substrate surface grooves. The width of the corrugated graphene region I, $L_g$, is to be determined later in the paper. When the spacing between the nanowires is small, the graphene tends to partially wrap around the surface envelope of the nanowires with a maximum corrugation amplitude $A_g$ which is smaller than the diameter of the nanowire, and a period that is equal to $L_{nw}$ (Fig. 1c). In this paper, we will quantitatively determine: (1) $L_g$ as a function of nanowire radius and interfacial bonding energies (graphene-substrate and graphene-nanowire) when $L_{nw} \gg L_g$; and (2) $A_g$ as a function of nanowire radius and interfacial bonding energies when $L_{nw}$ is comparable to or less than $L_g$.

To determine the regulated graphene morphology in the above two cases, we next delineate a computational model to quantitatively determine the interplay between two types of energies: (1) graphene-substrate and graphene-nanowire interaction energies and (2) system strain energy.

## III. COMPUTATIONAL MODEL

### A. INTERACTION ENERGIES

The interaction between mechanically-exfoliated graphene and the underlying substrate, as well as that between the graphene and the nanowires on the substrate are usually weak and can be characterized by van der Waals forces. The van der Waals force between an atomic pair of distance $r$ is represented by a Lennard–Jones pair potential $V_{LJ}(r) = 4\varepsilon(\sigma^{12}/r^{12} - \sigma^6/r^6)$, where $\sqrt[6]{2}\sigma$ is the equilibrium distance of the atomic pair and $\varepsilon$ is the bonding energy at the equilibrium distance. The graphene-substrate and the graphene-nanowire interaction energies are given by summing up all atomic pair interaction energies due to the van der Waals force between the carbon atoms in the graphene and the atoms in the nanowires and the substrate. The total interaction energy, denoted by $E_{\text{int}}$, including the interaction between a graphene of area $S$ and a substrate of volume $V_s$, as well as that between such a graphene and the underlying nanowires of volume $V_{nw}$, is then given by

$$E_{\text{int}} = \int_S \int_{V_s} V_{LJ}^s(r) \rho_s dV_s \rho_c dS + \int_S \int_{V_{nw}} V_{LJ}^{nw}(r) \rho_{nw} dV_{nw} \rho_c dS \qquad (1)$$

where $V_{LJ}^s$ and $V_{LJ}^{nw}$ are the Lennard–Jones pair potentials of the graphene-substrate interaction and the graphene-nanowire interaction, respectively, $\rho_c$ is the homogenized carbon atom area density of graphene that is related to the equilibrium carbon–carbon bond length $l$ by $\rho_c = 4/(3\sqrt{3}l^2)$, $\rho_s$ and $\rho_{nw}$ are the molecular density of the substrate and the nanowires, respectively, which can be derived from the molecular mass and mass density.

Since van de Waals interactions decay rapidly beyond equilibrium atomic pair distance, the multiple domain integral in Eq. (1) can be estimated by applying a cut-off distance for all atomic pair interactions. In all simulations reported in this paper, a sufficiently large cut-off distance of 3 nm is used, which leads to variations in the estimated value of $E_{int}$ less than 1%. We have established a Monte-Carlo numerical scheme to compute the multiple domain integrals in Eq. (1), which is briefly summarized as follows. For the $i^{th}$ carbon atom in the graphene, $n$ random locations are generated in the substrate portion within the cut-off distance from this carbon atom. The interaction energy between this carbon atom and the substrate is estimated by

$$E_i^s = (\rho_s V_s/n) \sum_{i=1}^{n} V_{LJ}^s(r_{ij}), \qquad (2)$$

where $r_{ij}$ is the distance between the $i^{th}$ carbon atom in the graphene and the $j^{th}$ random substrate location. Equation (2) is evaluated at $m$ equally spaced locations over the graphene of area $S$. The graphene–substrate interaction energy over this area can then be estimated by

$$E_{int}^s = (\rho_c S/m) \sum_{j=1}^{m} E_i^s = (\rho_c \rho_s SV_s/nm) \sum_{j=1}^{m} \sum_{i=1}^{n} V_{LJ}^s(r_{ij}). \qquad (3)$$

Following the similar scheme, the graphene-nanowire interaction energy can be estimated by

$$E_{int}^{nw} = (\rho_c \rho_{nw} SV_{nw}/nm) \sum_{j=1}^{m} \sum_{i=1}^{n} V_{LJ}^{nw}(r_{ij}). \qquad (4)$$

The total van de Waals interaction energy can be then estimated by

$$E_{\text{int}} = E_{\text{int}}^{s} + E_{\text{int}}^{nw}. \tag{5}$$

As *n* and *m* become sufficiently large, the estimated value of the interaction energy given by Eq. (5) converges to the theoretical value of the interaction energy defined by Eq. (1).

The above considerations of interaction energy is generally applicable to any given graphene morphology regulated by patterned nanowires on a substrate.

## B. STRAIN ENERGY OF THE SYSTEM

The strain energy in the system of graphene spontaneously regulated by nanowires patterned on a substrate surface results from the corrugating deformation of the graphene and the interaction-induced deformation of the substrate and the nanowires. When an ultrathin monolayer graphene partially conforms to rigid nanowires patterned on a rigid substrate (e.g., $SiO_2$ nanowires on a $SiO_2$ substrate), the resulting deformation of the substrate and the nanowires due to the weak graphene–substrate and graphene-nanowire interactions is expected to be negligible. Under such an assumption, the strain energy of the system is dominated by the strain energy of the graphene, which results from the out-of-plane corrugation and in-plane stretching of the graphene under nanowire/substrate regulation. The resulting strain energy in the substrate and the nanowires is thus not considered in this paper. Further discussion on the above assumption is to be further elaborated in Section V.

The deformation of the regulated graphene consists of both out-of-plane bending and in-plane stretching. Denoting the out-of-plane corrugation of the graphene by $w(x, y)$, the bending energy of the graphene is given by

$$E_b = \int_S D \left[ \frac{1}{2}\left(\frac{\partial^2 w}{\partial x^2} + \frac{\partial^2 w}{\partial y^2}\right)^2 - (1-\nu)\left(\frac{\partial^2 w}{\partial x^2}\frac{\partial^2 w}{\partial y^2} - \left(\frac{\partial^2 w}{\partial x \partial y}\right)^2\right)\right] dS, \quad (6)$$

where $D$ and $\nu$ are the bending rigidity and the Poisson's ratio of graphene, respectively.

The membrane energy of the graphene due to in-plane stretching is given by

$$E_s = \int_S \frac{Eh}{2(1-\nu^2)}\left[(\varepsilon_{xx} + \varepsilon_{yy})^2 + 2(1-\nu)(\varepsilon_{xy}^2 - \varepsilon_{xx}\varepsilon_{yy})\right] dS, \quad (7)$$

where $E$ and $h$ are the Young's Modulus and the thickness of graphene, respectively, $\varepsilon_{xx}$, $\varepsilon_{yy}$ and $\varepsilon_{xy}$ are the components of the in-plane membrane strain of the corrugated graphene.

The total strain energy of the graphene is then given by

$$E_g = E_b + E_s. \quad (8)$$

We next apply the above consideration to compute the graphene strain energy for the two cases of nanowire spacing described in Section II.

*Case 1: Graphene regulated by widely spaced nanowires on a substrate surface*

As discussed in Section II, when the spacing between the nanowires is large (i.e., $L_{nw} \gg L_g$), the graphene strain energy results from the deformation of the graphene portion in region I. The graphene portion in region II is undeformed and thus has no contribution to the strain energy. Given the symmetry of the configuration (Fig. 1b), the out-of-plane corrugation of graphene in region I, $w(x)$, is taken to be described by a cubic polynomial of $x$, whose coefficients can be determined by the following boundary conditions: $w = 0$ and $dw/dx = 0$ at $x = 0$, and $w = -d_{nw}$ and $dw/dx = 0$ at $x = L_g/2$. This consideration leads to

$$w(x) = 4d_{nw}\left[4\left(\frac{x}{L_g}\right)^3 - 3\left(\frac{x}{L_g}\right)^2\right]. \tag{9}$$

We next determine the in-plane membrane strain of the graphene. Given the symmetry of the configuration, the membrane strain components $\varepsilon_{yy}$ and $\varepsilon_{xy}$ are taken to be zero. At the equilibrium morphology, in-plane shear stress acting on the graphene vanishes, which leads to a constant non-zero membrane strain $\varepsilon_{xx}$ in the graphene portion in region I. That is,

$$\varepsilon_{xx} = \frac{du}{dx} + \frac{1}{2}\left(\frac{dw}{dx}\right)^2 = \text{constant}. \tag{10}$$

where $u(x)$ is in-plane displacement of graphene in $x$ direction. The symmetric configuration requires the boundary conditions of $u(0) = u(L_g/2) = 0$. The above consideration leads to

$$\varepsilon_{xx} = \frac{12d_{nw}^2}{5L_g^2}. \tag{11}$$

Substituting Eqs. (9) and (11) into Eqs. (6) and (7), respectively, the strain energy of the graphene is given by

$$E_g = E_b + E_s = \frac{48Dd_{nw}^2}{L_g^3} + \frac{36Ehd_{nw}^4}{25L_g^3(1-v^2)}. \tag{12}$$

*Case 2: Graphene regulated by densely spaced nanowires on a substrate surface*

When the spacing between the nanowires is small, the graphene partially conforms to the nanowires. The out-of-plane corrugation and the in-plane stretching of the graphene can be determined by a similar approach described above but with different boundary conditions.

For the out-of-plane corrugation of the graphene $w(x)$, the boundary conditions in this case are $w = 0$ and $dw/dx = 0$ at $x = 0$, and $w = -A_g$ and $dw/dx = 0$ at $x = L_{nw}/2$, where $A_g$ is the maximum amplitude of the graphene corrugation that remains to be determined. These considerations lead to

$$w(x) = 4A_g \left( 4\left(\frac{x}{L_{nw}}\right)^3 - 3\left(\frac{x}{L_{nw}}\right)^2 \right). \tag{13}$$

For the in-plane displacement $u(x)$ of the graphene, the symmetric configuration requires the boundary conditions of $u(0) = u(L_{nw}/2) = 0$. These considerations lead to

$$\varepsilon_{xx} = \frac{12A_g^2}{5L_{nw}^2}. \tag{14}$$

Accordingly, the strain energy of the graphene in this case is given by

$$E_g = E_b + E_s = \frac{48DA_g^2}{L_{nw}^3} + \frac{36EhA_g^4}{25L_{nw}^3(1-v^2)}. \tag{15}$$

## C. REGULATED MORPHOLOGY OF GRAPHENE

The computational models described in Sections III.A and III.B are used to compute the total system free energy ($E_{int} + E_g$) of the following two simulation cases.

*Case 1: Graphene regulated by widely spaced nanowires on a substrate surface*

In this case, the graphene portion near the nanowires corrugates and wraps around each nanowire, while the rest portion of graphene remains flat on the substrate surface. As shown in Eq. (12), the graphene strain energy $E_g$ monotonically decreases as the width of corrugated graphene portion $L_g$ increases. On the other hand, due to the nature of van der Waals interaction, the interaction energy $E_{int}$ minimizes at a finite value of $L_g$. As a result, there exists a minimum value of the total system free energy ($E_{int} + E_g$), where $L_g$ (i.e., the width) and $d_{nw}$ (i.e., the amplitude) define the equilibrium morphology of the graphene regulated by widely spaced nanowires on a substrate surface.

*Case 2: Graphene regulated by densely spaced nanowires on a substrate surface*

In this case, the graphene partially conforms to the nanowires. As shown in Eq. (15), the graphene strain energy $E_g$ monotonically increases as the maximum amplitude of the graphene corrugation $A_g$ increases. On the other hand, due to the nature of van der Waals interaction, the interaction energy $E_{int}$ minimizes at a finite value of $A_g$. As a result, there exists a minimum value of the total system free energy ($E_{int} + E_g$), where $L_{nw}$ (i.e., the period) and $A_g$ (i.e., the amplitude) define the equilibrium morphology of the graphene regulated by densely spaced nanowires on a substrate surface.

The energy calculation and minimization are carried out by running a customized code on a high performance computer cluster. In all simulations, we use $D = 1.41\ eV$, $E = 1\ TPa$, $h = 0.34 nm$, $\nu = 0.4$, $l = 0.142\ nm$, $\rho_s = \rho_w = 2.20 \times 10^{28} / m^3$ and $\sigma = 0.353\ nm$. These parameters are representative of a graphene/SiO$_2$ nanowire/SiO$_2$ substrate material system.[31,32] Various values of $d_{nw}$, $L_{nw}$ and $\varepsilon$ are used to study the effects of nanowire size and spacing as well as the interfacial bonding energy on the regulated graphene morphology.

## IV. RESULTS AND DISCUSSION

### A. GRAPHENE REGULATED BY WIDELY SPACED NANOWIRES ON A SUBSTRATE SURFACE

Figure 2a plots the equilibrium width of the corrugated graphene region I, $L_g$, as a function of the nanowire diameter $d_{nw}$ for various values of $D/\varepsilon$. For a given interfacial bonding energy (i.e., a given value of $D/\varepsilon$), $L_g$ increases as the nanowire diameter $d_{nw}$ increases in an approximately linear manner, as indicated by the straight fitting lines in Fig. 2. In other words, the width of the corrugated graphene region is generally linearly proportional to the nanowire size. For a given nanowire diameter $d_{nw}$, $L_g$ decreases as the interfacial bonding energy between graphene and nanowire/substrate increases (i.e., smaller value of $D/\varepsilon$). That is, a stronger interfacial bonding results in a narrower region of corrugated graphene. The effect of interfacial bonding energy on graphene morphology is further elucidated by the $L_g$ vs. $D/\varepsilon$ curves in Fig. 2b for various values of $d_{nw}$. Emerging from Fig. 2b is an apparent power law dependence of $L_g$ on $D/\varepsilon$. Combined with the linear dependence of $L_g$ on $d_{nw}$ that is evident in Fig. 2a, the correlation between the width of corrugated graphene region and the nanowire size as well as the interfacial bonding energy can be described by

$$\frac{L_g}{d_{nw}} \cong 3.84 \left(\frac{D}{\varepsilon}\right)^{\frac{1}{4}}. \tag{16}$$

Together with Eq. (9), Eq. (16) offers a rule-of-thumb estimate of the graphene morphology regulated by widely spaced nanowires on a substrate surface, agreeing with the full-scale simulation results within 5%.

**B. GRAPHENE REGULATED BY DENSELY SPACED NANOWIRES ON A SUBSTRATE SURFACE**

Figure 3a plots the amplitude of graphene corrugation normalized by the nanowire diameter $A_g/d_{nw}$ as a function of the spacing between nanowires $L_{nw}$ for two nanowire diameters $d_{nw} = 2.0nm$ and $3.2nm$. Here, $D/\varepsilon = 300$. For a given nanowire size (i.e., $d_{nw}$), if the spacing between nanowires $L_{nw}$ is large, $A_g/d_{nw}$ tends to one. In other words, the graphene can fully wrap around the nanowires, leading to a corrugated morphology that can be described by $w(x) = 4d_{nw}(4(x/L_{nw})^3 - 3(x/L_{nw})^2)$. By contrast, if the spacing between nanowires $L_{nw}$ is small, $A_g/d_{nw}$ approaches zero. That is, the graphene is nearly flat and does not conform to the patterned nanowires. Such a trend can be understood as follows. For a given nanowire size, if the spacing between nanowires is small, conforming to each nanowire results in a significant increase in the graphene strain energy (note that $E_g \propto 1/L_{nw}^3$ in Eq. (15)). Consequently, $A_g$ tends to zero. On the other hand, if $L_{nw}$ is large, the resulting graphene strain energy is limited even when $A_g/d_{nw}$ tends to one. Consequently, the graphene tends to closely follow the surface envelope of the patterned nanowires.

A significant feature shown in Fig. 3a is the sharp transition in the equilibrium amplitude of the graphene corrugation as the nanowire spacing varies. For example, for $d_{nw} = 3.2nm$, $A_g/d_{nw}$ raises abruptly from 0.1 to 1.0, as $L_{nw}$ varies slightly from 52.8$nm$ to 54.4$nm$. In other words, the graphene morphology snaps between two distinct states (Fig. 3b): (1) closely conforming to the envelope of the nanowires patterned on a substrate surface and (2) nearly remaining flat on the nanowires patterned on a substrate surface, when the spacing of nanowires reaches a critical value, $L_{nw}^{cr}$. Further comparison shows that the critical nanowire spacing is approximately equal to the width of corrugated graphene region $L_g$ determined in Section 4.1 (e.g., in Fig. 2 or Eq. (16)). For example, Eq. (16) gives $L_g = 51.1nm$ for $d_{nw} = 3.2nm$ and $D/\varepsilon = 300$, which agrees well with the critical nanowire spacing defined in Fig. 3. Such a snap-through instability of graphene morphology is similar to that of graphene morphology regulated by the underlying substrate surface with engineered nanoscale patterns (e.g., surface grooves, herringbone or checkerboard wrinkles). It has been shown that the snap-through instability of graphene morphology results from a double-well profile of total free energy of the system (i.e., $E_g + E_{int}$) as a function of the amplitude of the graphene morphology (e.g., $A_g$) at a critical value of surface feature size (e.g., nanowire spacing, surface groove roughness, etc.).[22,23] For example, the simulation results in this paper show that ($E_g + E_{int}$) minimizes at two values of $A_g$ when $L_{nw} = L_{nw}^{cr}$.

Emerging from the results in Sections 4.1 and 4.2 is a coherent understanding of the graphene morphology regulated by nanowires patterned in parallel on a substrate surface: for a given nanowire size (e.g., $d_{nw}$) and graphene/nanowire/substrate interfacial bonding energy (e.g., $D/\varepsilon$), there exists a critical nanowire spacing $L_{nw}^{cr}$, which can be estimated by $L_{nw}^{cr} \cong 3.84 d_{nw} (D/\varepsilon)^{1/4}$. If the nanowire spacing $L_{nw}$ is greater than $L_{nw}^{cr}$, graphene can corrugate to wrap around the nanowires, with a maximum amplitude equal to the nanowire diameter and in a region of width equal to $L_{nw}^{cr}$. The morphology of the corrugated portion of graphene can then be described by

$$w(x) = \begin{cases} 4d_{nw}(4(x/L_{nw}^{cr})^3 - 3(x/L_{nw}^{cr})^2) & 0 < x < L_{nw}^{cr} \\ 0 & L_{nw}^{cr} < x < L_{nw} \end{cases}. \quad (17)$$

By contrast, if the nanowire spacing $L_{nw}$ is smaller than $L_{nw}^{cr}$, graphene remains near flat on the patterned nanowires. When $L_{nw} = L_{nw}^{cr}$, the regulated graphene morphology snaps between the above two distinct states.

The above understanding also implies that, besides the nanowire spacing, the interfacial bonding energies can influence the graphene morphology. Figure 4 further shows the effect of $D/\varepsilon$ on the normalized amplitude of graphene corrugation $A_g/d_{nw}$ for various values of $L_{nw}$. Here, $d_{nw} = 3.2 nm$. For a given nanowire spacing $L_{nw}$, if the interfacial bonding is strong (i.e., small $D/\varepsilon$), $A_g/d_{nw}$ tends to one (i.e., graphene wraps around nanowires); if the interfacial bonding is weak (i.e., large $D/\varepsilon$), $A_g/d_{nw}$ approaches zero (graphene does not conform to the nanowires). There also exists a snap-through instability at which the graphene morphology switches abruptly between the abovementioned two distinct states when $D/\varepsilon$ reaches a critical value, for a given value of $L_{nw}$. For example, such a critical value of $D/\varepsilon$ is about 15, 100 and 400 for $L_{nw}$=24*nm*, 40*nm* and 56*nm*, respectively. Substituting these critical values of $D/\varepsilon$ into Eq. (16) yields $L_{nw}$ = 24.2*nm*, 38.9*nm* and 55.0*nm*, respectively, further demonstrating quite well agreement with the simulation results.

## V. CONCLUDING REMARKS

In this paper, we determine the graphene morphology regulated by nanowires patterned in parallel on a substrate surface through energy minimization. The equilibrium graphene morphology is governed by the interplay between the corrugation-induced strain energy of the graphene and the interaction energy between the graphene and the underlying nanowires and substrate. The graphene strain energy consists of the contribution from both out-of-plane bending and in-plane stretching, which are derived from nonlinear plate theory. The interaction energy is characterized by the sum of all atomic pair potential between the graphene carbon atoms and the substrate atoms/molecules, which is computed through a Monte Carlo type numerical scheme. The major conclusions emerging from the modeling results are summarized as follows:

- The graphene morphology on nanowires evenly patterned in parallel on a substrate surface can be regulated by the nanowire size, nanowire spacing, and interfacial bonding energy between the graphene and the nanowire and that between the graphene and the substrate surface.

- For a given nanowire size and a given interfacial bonding energy, there exists a critical nanowire spacing, greater than which the graphene can conform to the surface envelope of the patterned nanowires, while smaller than which the graphene remains nearly flat on the nanowires. The graphene morphology snaps between these two distinct states at the critical nanowire spacing. The conforming morphology of the graphene can be quantitatively determined.

- For a give nanowire size and spacing, there exists a critical interfacial bonding energy, higher than which the graphene can conform to the surface envelope of the patterned nanowires, while lower than which the graphene remains nearly flat on the nanowires. The snap-through instability of graphene morphology also exists at this critical interfacial bonding energy.

- The abovementioned critical nanowire spacing, critical interfacial bonding energy and the nanowire size can be correlated by $L_{nw}^{cr}/d_{nw} \cong 3.84(D/\varepsilon)^{1/4}$, a rule-of-thumb estimate that agrees quite well with the full-scale simulation results.

In the present model, we consider graphene morphology regulated by rigid nanowires (e.g., $SiO_2$) patterned on a rigid substrate (e.g., $SiO_2$). The graphene corrugation induced deformation in the nanowires and the substrate is expected to be negligible. Recent experimental progress enables transfer printing graphene onto a wide variety of substrate surfaces[9,33-35] (e.g., polymers or elastomers). The graphene-substrate interaction may result in appreciable deformation of the underlying compliant polymer or elastomer substrate. To extend the results from the present study to such a case, the strain energy of the substrate needs to be considered in the energy minimization. The present model also assumes the weak graphene-nanowire/substrate interaction. In practice, it is possible to have chemical bonding or pinnings[36-38] between the graphene and the substrate, the nanowire surface can also be functionalized to facilitate chemical bonding with the graphene, both of which lead to an enhanced interfacial bonding. Recent experiments report the blister morphology in thin graphene sheet due to the intercalation of gold nanoparticles between the graphene and the underlying silicon wafer.[39] The size of such blisters is shown to be correlated with the graphene-silicon adhesion and the nanoparticle diameter. The energetic framework and the numerical strategy reported in the present paper still holds and can be readily adapted to determine the graphene morphology regulated by nanoparticles intercalated along the graphene-substrate interface. The research in this regard will be reported elsewhere.

In summary, we further extended our earlier energetic research framework to quantitatively determine the graphene morphology regulated by nanowires of diameters of one to a few nanometers patterned in parallel on a substrate surface. The critical physical and geometric parameters that define the characteristics of the regulated graphene morphology can be correlated through a rule-of-thumb formula in quite well agreement with full-scale simulation results. Such a formula can serve as a first order guideline in nano-structural design to achieve certain desired graphene morphology via nanowire regulation. The results from the present study further demonstrate the potential to regulate the graphene morphology with ultra fine resolution against its intrinsic randomness via patterned nanowires with diameters approaching the regime of one to a few nanometers.[28]

**ACKNOWLEDGMENT:**

This work is supported by the Minta-Martin Foundation, a UMD General Research Board summer research award to T. L., National Science Foundation (Grant No. 0928278). Z.Z. also thanks the support of the A. J. Clark Fellowship.

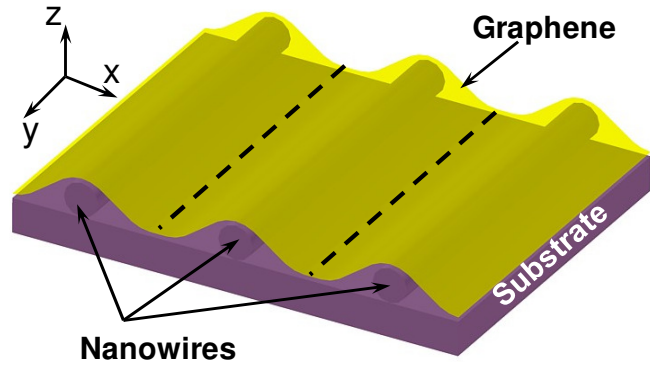

(a)

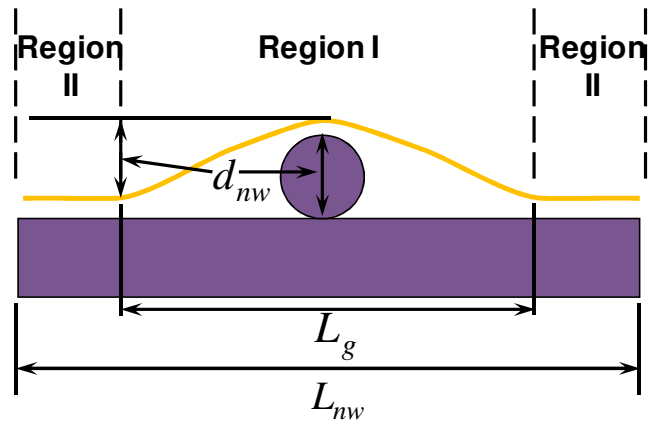

(b)

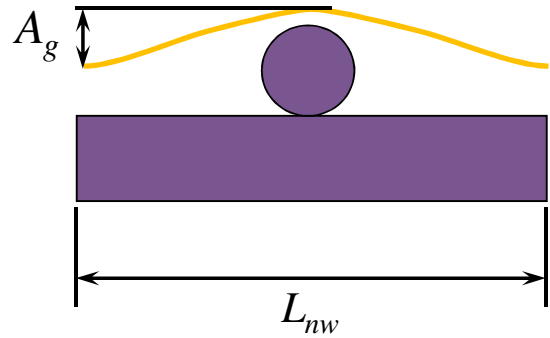

(c)

Fig. 1. (Color online) (a) Schematic of a blanket graphene regulated by nanowires patterned in parallel on a substrate surface. The graphene between the two dashed lines and the underlying

nanowire and the substrate are modeled (e.g., in (b) and (c)) due to the configuration symmetry. (b) and (c) depict two limiting cases of the regulated graphene morphology. (b) If the nanowire spacing $L_{nw}$ is large, graphene corrugates to wrap around the nanowire in a region of width $L_g$ (Region I) and remains flat on the substrate in Region II. The amplitude of the graphene corrugation is equal to the nanowire diameter $d_{nw}$. (c) If the nanowire spacing $L_{nw}$ is small, graphene corrugates to partially conform to the envelope of the nanowire surfaces, with a period of $L_{nw}$ and an amplitude $A_g$ which is much smaller than $d_{nw}$.

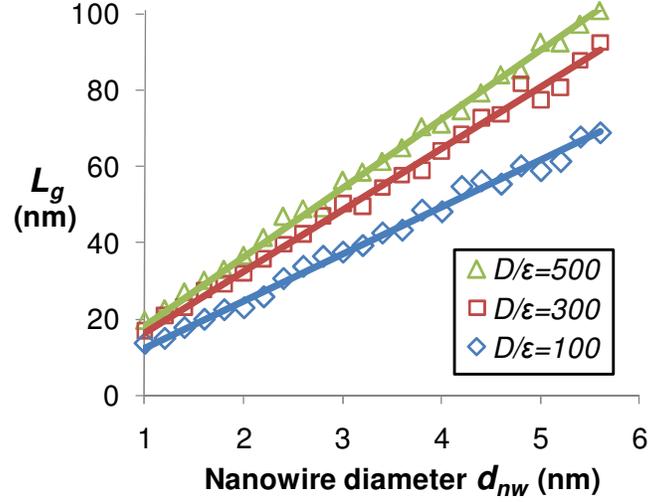

(a)

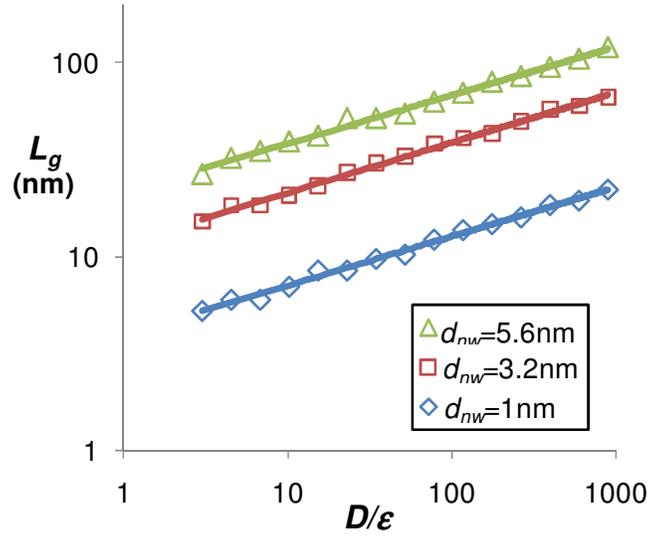

(b)

Fig. 2. (Color online) (a) The equilibrium width of the corrugated graphene region I, $L_g$, as a function of the nanowire diameter $d_{nw}$ for various values of $D/\varepsilon$. The straight fitting lines denotes the linear dependence of $L_g$ on $d_{nw}$. (b) $L_g$ as a function of $D/\varepsilon$ for various values of $d_{nw}$. The straight fitting lines denotes the power law dependence of $L_g$ on $D/\varepsilon$ (note the logarithmic scales of both axes).

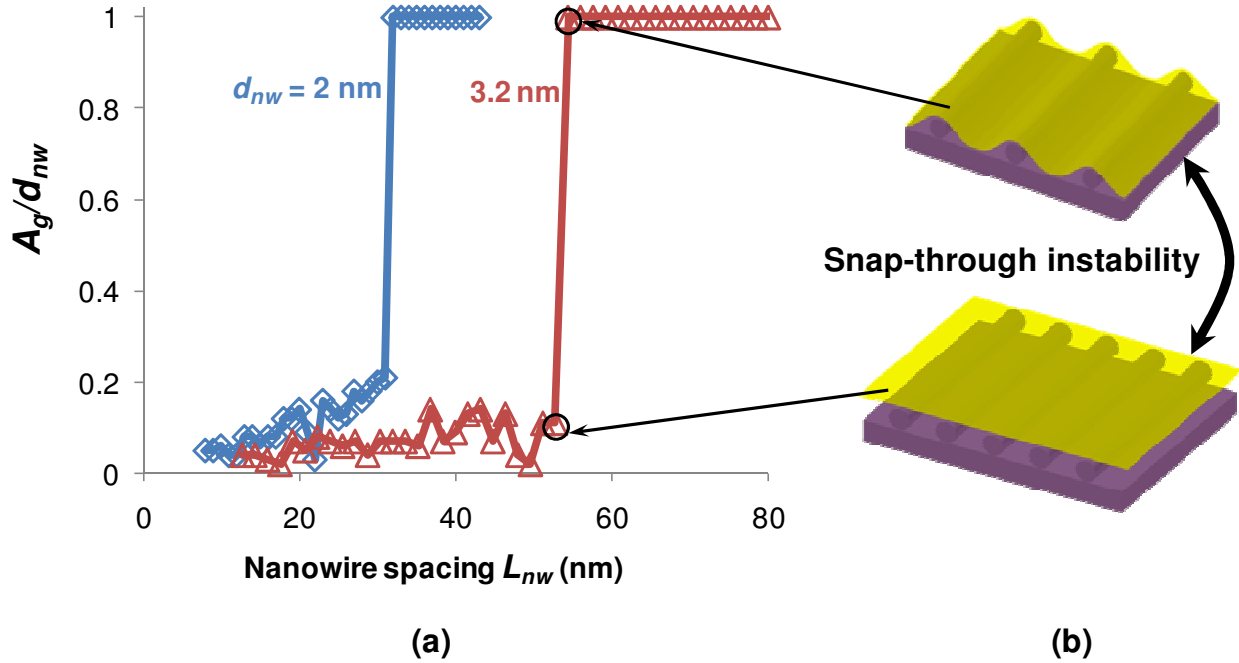

Fig. 3. (Color online) (a) The amplitude of graphene corrugation normalized by the nanowire diameter $A_g/d_{nw}$ as a function of nanowire spacing $L_{nw}$ for two nanowire diameters $d_{nw} = 2.0 nm$ and $3.2 nm$. Here, $D/\varepsilon = 300$. As the nanowire spacing approaches a critical value $L_{nw}^{cr}$, the graphene morphology snaps between two distinct states: (1) closely conforming to the envelope of the nanowires patterned on a substrate surface and (2) nearly remaining flat on the nanowires patterned on a substrate surface, as illustrated in (b).

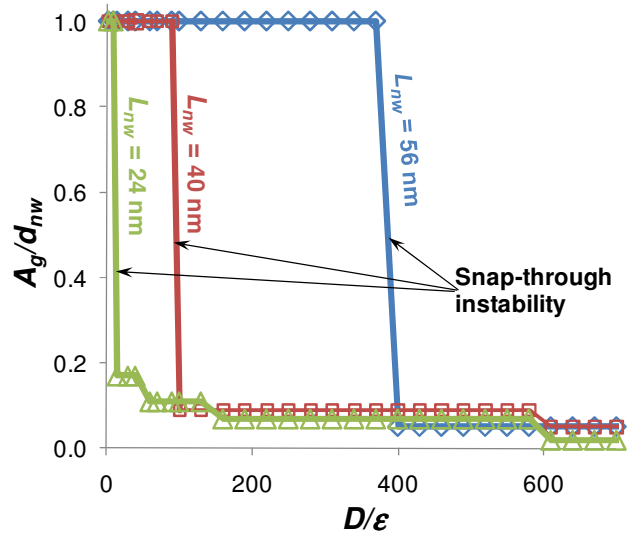

Fig. 4. (Color online) $A_g/d_{nw}$ as a function of $D/\varepsilon$ for various values of $L_{nw}$. Here, $d_{nw} = 3.2 nm$. Note the snap-through instability similar with that in Fig. (3) as $D/\varepsilon$ reaches a critical value for a given value of $L_{nw}$.